\documentclass[12pt]{article}
\usepackage{amsmath}
\usepackage[margin=2.0cm]{geometry}
\usepackage{epsfig}
\usepackage{amsfonts}
\usepackage{fancyhdr}
\usepackage{setspace}
\usepackage{enumitem}
\usepackage[font={sf, sl}, labelfont={sf,sl}, margin=1cm]{caption}
\usepackage{graphicx}
\usepackage{subcaption}
\usepackage{bm}
\usepackage{cancel}
\usepackage{wasysym}
\usepackage{amssymb}
\usepackage{wrapfig}
\usepackage{textcomp}
\usepackage{arydshln}
\usepackage{tikz}
\usepackage{authblk}
\usepackage{framed}
\usepackage{titlesec}
\usepackage{hyperref}
\usepackage{cite}
\usepackage[page,toc,titletoc,title]{appendix}
\usetikzlibrary{arrows}
\newcommand{\authorname}{Nitin Gupta} 
\titleformat{\section}
{\normalfont\bfseries\scshape}{\thesection}{1em}{}

\titleformat{\subsection}
{\normalfont\bfseries\scshape}{\thesubsection}{1em}{}

\titleformat{\subsubsection}
{\normalfont\slshape}{\thesubsubsection}{1em}{}

\titleformat{\title}{\normalfont\bfseries}{\thesection}{1em}{}

\title{\vspace{-3em} \large Synthetic space bricks from lunar and martian regolith via sintering}
\author[1]{\normalsize \authorname}
\author[1]{\normalsize Vineet Dawara}
\author[1]{\normalsize Aloke Kumar}
\author[1]{\normalsize Koushik Viswanathan \thanks{koushik@iisc.ac.in}}
\affil[1]{\normalsize \textsl{Department of Mechanical Engineering, Indian Institute of Science, Bangalore}}
\date{\normalsize \textsc{\today}}
\begin{document}	
\maketitle
\thispagestyle{plain}	
\hrulefill

\begin{abstract}
The prospect of establishing extra-terrestrial habitats using in situ resource utilization (ISRU) constitutes a long-term goal of multiple space agencies around the world. In this work, we investigate sintering as a potential route for making building blocks---termed synthetic space bricks---using \emph{in situ} regolith material. By systematically investigating sintering parameters using a numerical lattice model, coupled with experimental observations and post sintering characterization, we propose a process protocol for two lunar---lunar highland simulant (LHS) and lunar mare dust simulant (LMS)---and one martian (martian global simulant, MGS) simulants. The resulting bricks demonstrate compressive strengths of upto 45 MPa under uniaxial loading, depending on the simulant used. These strengths are much greater than those typically mandated for structural applications under reduced gravity. We infer microscale sintering mechanisms at the individual particle level indirectly, by measuring temporal evolution exponents of sample dimensions during sintering. For all three simulants, volume diffusion appears to be the primary mechanism for particle coalescence. Our results clearly make a strong case for the use of sintering as a potentially scalable method for consolidating regolith into brick-like structures for load-bearing applications in extra-terrestrial settings.
\end{abstract}

\section{Introduction}
The prospect of establishing extra-terrestrial habitats continues to remain a long-term goal of multiple space agencies around the world \cite{scoville2022artemis}. This field has traditionally encompassed a wide variety of disciplines ranging from biological effects on humans \cite{rambaut1975calcium} to geology and civil engineering \cite{schmitt1973apollo}. The possibility of developing extra-terrestrial structures that can sustain both extreme environments and be built with minimal raw materials has drawn significant recent interest \cite{labeaga2017additive, stenzel2018sustainable}. The obvious option of exploiting intrinsic features on the martian or lunar surface, for instance lava tubes \cite{angelis2002lunar, theinat2020lunar}, is fraught with fundamental difficulties such as potential structural collapse and the occurrence of surface fissures. The alternative option of constructing settlements with material sourced from earth is expected to incur significant costs, rendering it practically unviable.

A recently emerging research paradigm---\emph{in situ} resource utilization or ISRU---attempts to address this problem by developing technologies for exploiting resources, primarily regolith and solar energy, available on the martian or lunar surface \cite{crawford2014lunar, benaroya2016turning, cockell2012uninhabited}. To help seed research in this direction, several artificial regolith simulants have been developed to mimic martian or lunar soil corresponding to various known locations on mars and the moon, respectively, based on corresponding spectroscopy and/or particle morphology data \cite{venugopal2020invention, mckay1994jsc,  fackrell2021development, isachenkov2022characterization, cannon2019mars, engelschion2020eac, li2009nao, ryu2018development, kanamori1998properties}. These regolith simulants can be consolidated to form load bearing structures for habitat applications using a range of strategies, from exploiting biological routes \cite{roberts2021blood, roedel2014protein, dikshit2021space, dikshit2022microbial} to employing concrete-mimicking processes \cite{cullingford1992lunar, hatanaka2004hydration, snehal2023development} and polymer binders with sintering \cite{hintze2009lunar, han2022sintering, zocca2020investigation}.

Among these consolidation techniques, sintering-based routes have so far yielded the most promising results when final part strength is the primary requirement. Various sintering protocols that have hitherto been proposed include microwave \cite{gholami2022hybrid, taylor2005microwave, kim2021microstructural}, spark plasma \cite{zhang2021spark}, cold-sintering \cite{karacasulu2023cold} and solar radiation \cite{ghosh2016solar, fateri2019solar, meurisse2018solar}. The effects of various process parameters on resulting consolidate strength have also been systematically investigated at the macroscale, including the role of sintering temperature \cite{warren2022effect, dou2019sintering}, porosity \cite{gualtieri2015compressive, indyk2017structural}, initial mineral compositions \cite{meurisse2017influence, osio2021sintering} and the presence of glass-like phases \cite{zheng2022microstructure}. A summary of these sintering/consolidation techniques and the resulting average compressive strength $(\sigma_{comp, avg})$ is presented in table \ref{sintering_LR}. 
\begin{table}[ht!]
	\fontsize{11pt}{11pt}\selectfont
	\centering
	\begin{tabular}{|c|c|c|c|c|c|} \hline
		\textbf{Consolidation technique} & \textbf{Temperature }& \textbf{Simulants } & \textbf{$\sigma_{comp, avg}$ } & \textbf{Ref.}\\ 
		 & \textbf{($^\circ$C) }& \textbf{used} & \textbf{ (MPa)} & \\ [0.5ex] \hline
		 Sintering & 1000-1100 & HUST-1, CAS-1 & 68 & \cite{han2022sintering} \\ [0.5ex]  \hline
		Sintering \& melting & 1000-1300   & JSC-2A & 31 &  \cite{zocca2020investigation} \\ [0.5ex]  \hline
		Hybrid microwave sintering & 1075-1125 & FJS-1 & 45& \cite{gholami2022hybrid} \\ [0.5ex]  \hline
		Microwave sintering & 1200-1500 &  MLS-1, JSC-1& -  & \cite{taylor2005microwave}\\ [0.5ex] \hline
		Microwave sintering & 1120 & KLS-1 & 37 &  \cite{kim2021microstructural} \\ [0.5ex] \hline
		Spark Plasma sintering & 1050 & FJS-1 & 220 &  \cite{zhang2021spark} \\ [0.5ex]  \hline
		Cold sintering & 250 & MGS-1 & 45 &  \cite{karacasulu2023cold}\\ [0.5ex]  \hline
		Sintering  & 1100-1200 & MGS-1, LMS-1 & 22-25 &   \cite{warren2022effect} \\[0.5ex] \hline 		
		Digital light  & 1100-1150 & CLRS-2  & 56 - 312  & \cite{dou2019sintering} \\
		processing \& sintering	&  &  &  &   \\ [0.5ex]  \hline
		Sintering & 1200 &  JSC-1 A, JSC-1AF,  & 103 - 232 & \cite{gualtieri2015compressive} \\
		&  & and JSC- 1AC   &  & \\ \hline
		Sintering & 1120 & JSC-1A  & 84.6 - 218.8  & \cite{indyk2017structural} \\ [0.5ex] \hline		
		Sintering & 1070-1125 & JSC-1, DNA, MLS-1  & 98 - 152 &  \cite{fischer2018situ, meurisse2017influence} \\ [0.5ex]  \hline
		Additive Manufacturing  & 1200 & LHS-1 & 20 &  \cite{osio2021sintering} \\ 
		\& sintering & & & & \\ [0.5ex] \hline
		Laser assisted sintering & 1400 & HIT-LRS & 68 &   \cite{zheng2022microstructure} \\ [0.5ex] \hline
		Solar Sintering &  & JSC-2A  & 2.49 & \cite{imhof2017advancing} \\
		and 3D printing &  &   &  & \\ [0.5ex]  \hline
		Extrusion \& sintering & 1050 & JSC-1A  & 20 & \cite{taylor2018sintering} \\ [0.5ex]  \hline
		Additive manufacturing & 1000 & EAC-1 & 5.4 &  \cite{altun2021additive} \\
		 \& sintering &  &  &  &  \\ [0.5ex]  \hline
		Brazing of SiC & 1400 & LRS, MRS & 21 - 27 &  \cite{zheng2022microstructure} \\ [0.5ex]  \hline
		Laser assisted sintering & 1000-1100 & Quartz sand &  $\sigma_{tensile}$ = 9.28 & \cite{zhao2022development} \\ [0.5ex]  \hline 
		Electric current assisted  & 700  & JSC-1A & 50 &  \cite{phuah2020ceramic} \\ 
		sintering (ECAS)& & & & \\ [0.5ex]  \hline
	\end{tabular}
	\caption{List of various sintering strategies used to consoildate regolith simulants, with reported average mechanical strength under uniaxial compression.}
	\label{sintering_LR}
\end{table}	

On the microscale, mechanisms governing particle consolidation during the sintering process are expected to be independent of the energy source utilized and largely determined by the thermal fields thereby produced \cite{kang2004sintering}. In this context, the specialized sintering techniques introduced allow for varying degrees of thermal control. Yet they cannot compare with furnace-based sintering, perhaps one of the oldest methods for producing ceramics. Here spatially uniform temperatures are \emph{a priori} the norm, given the lack of directionality of a focused energy source. Consequently parts produced via furnace sintering are expected to have uniform, isotropic properties; this process is also inherently scalable, as has been amply demonstrated on earth. An additional advantage of furnace sintering is that it enables more systematic study of the role of microscopic mechanisms operative at much smaller length scales without intervening spatial inhomogeneity effects.

The primary objectives of the present work are threefold---the first is to develop a scalable experimental protocol for making consolidated regolith-based bricks on both the lunar and martian surfaces by using a polymer-based binder and furnace sintering. To this end, we work with two lunar simulants---lumar highland simulant (LHS) and lunar mare dust simulant (LMS)---and one martian simulant (martian global simulant, MGS). The second is to evaluate the microscopic mechanisms based on the kinetics of the sintering process by taking recourse to classical results in the ceramics literature. The third and final task is to correlate the established processing protocol and operative microscopic mechanisms with final part strength. Our manuscript is organized as follows. We first discuss the strategy of brick manufacturing in Sec.~\ref{brick_fab}, and corresponding post-consolidation characterization (Sec.~\ref{charac}). A simple numerical model is presented to determine heating parameters needed for ensuring spatial homogeneity (Sec.~\ref{sec:thermal}). The primary results are described in Sec.\ref{results}, beginning with numerical estimation of minimum soaking time $t_s$ required (Sec.~\ref{simulation}), followed by evaluation of mechanical properties (Sec.~\ref{CT}) and sintering mechanisms (Sec.~\ref{mech}). We present a discussion of our results and provide concluding results in Sec.~\ref{sec:conclusions}.

\section{Materials and Methods}
\subsection{Protocol for single brick production via sintering}\label{brick_fab}
\begin{figure}[ht!]
	\centering	
	\includegraphics[width=0.85\textwidth]{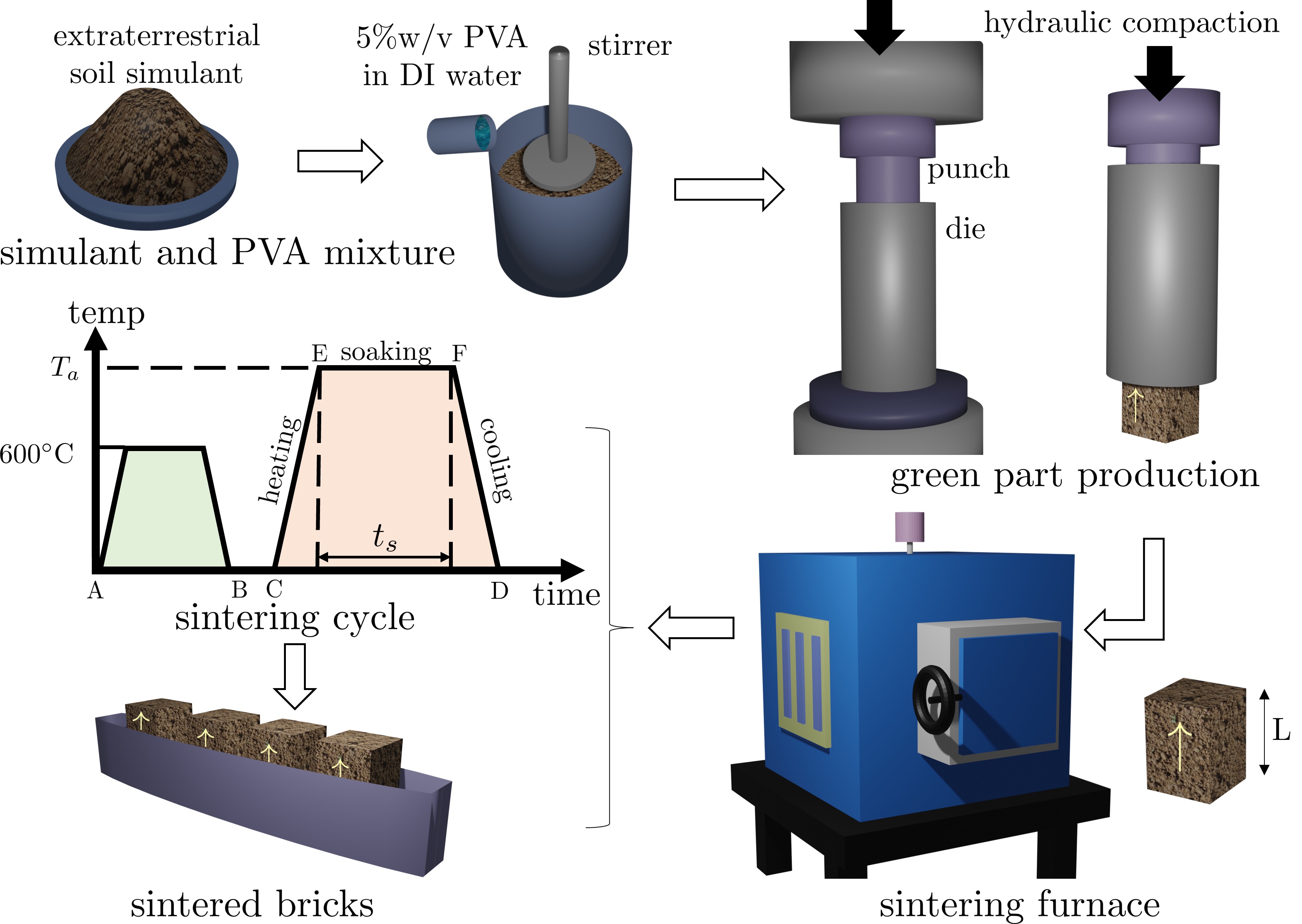}
	\caption{Schematic representation of the sintering process. Mixing of 5\% PVA in 100 ml of DI water mixture extra-terrestrial soil simulant within the ratio of 0.15:1 w/w. The mixture is poured into a die and compacted using a hydraulic press, followed by sintering in the furnace. The sintering cycle represents the preparation of the green part (heating to 600$^\circ$C for 1 hr) followed by the brown part (heating to $T_a$ for $t_s$ minutes).}		
	\label{schematic}
      \end{figure}

We use three types of soil simulants for our experiments---MGS (Martian global simulant), LHS (Lunar highland simulant), and LMS (Lunar mare dust simulant), procured from Exolith lab, Florida, USA \cite{isachenkov2022characterization, cannon2019mars}; details of these simulants are provided in Fig.~S2 and Table~1 of supplementary material. The experimental procedure for producing sintered parts is summarized in Fig.~\ref{schematic}. A PVA solution was prepared by mixing 5g of PVA (polyvinyl alcohol) powder (molecular weight 1,15,000 from Loba chemie pvt ltd.) with 100 ml of DI water and stirring at 90$^\circ$C for 1 hr, followed by stirring at room temperature for 10-12hr. This solution (15 ml) was then thoroughly mixed with 100g of simulant and the mixture die cast in the form of a cubical block of approximately $\sim 18\times 18\times 18$ mm$^3$ using a hydraulic press with 280-300 MPa compaction pressure. The resulting compacted sample weighed around 14 grams; it was then heated in a muffle furnace (Delta Power systems) for sintering.

Stages of the sintering cycle are shown in the form of a temperature vs. time curve, see Fig.~\ref{schematic} (bottom left). In the first step, as-cast samples were heated for 1 hour at 600$^\circ$C with a slow temperature ramp-up and ramp-down. This stage removes any volatile matter from the bricks, along with the PVA binder used to make the compacted part. The furnace was then brought back to room temperature at the time marked B. The dimensions of the sample were measured at this stage and are henceforth referred to as the intiial dimensions, denoted $L_i$. Correspondingly, the sample weight at the end of this stage typically reduced by $\sim$13-14\% . At time B, the sample is referred to as a green part.

The green part was then subject to the next heating stage (C to D) upto a peak temperature of $T_a = 1150^\circ$C with heating rate $c = 5^\circ$C/min for time $t_f$ (point C to E). Following this, the samples were soaked for different times ($t_s$) ranging from 10 minutes to 480 minutes (point E to F), and then cooled (point F to D) at a rate of 4$^\circ$C/min to obtain final brown parts, which we term \lq synthetic space bricks\rq . The final dimensions $L_f$ were measured post recovery to room temperature; the sample weight was typically found to reduce by $\sim$4\% at this stage, compared to the green part. 

\subsection{Post-sintering characterization}\label{charac}	
Internal porosity of sintered bricks was evaluated using mercury intrusion porosimetry (Pore Master), henceforth referred to as MIP. This technique uses high pressure (35000 psi) to drive mercury into the pore spaces in order to determine the pore size distribution ranging from sub-micrometer pores to a few hundred micrometers. Compressive strength measurements were performed using quasi-static displacement-controlled unaxial compression on a universal testing machine (Instron-5697) with a 30 kN capacity load cell and loading rate of 0.5 mm/min. Post-sintering microstructural examination was carried out using FE-SEM (field emission-scanning electron microscopy)(Carl Zeiss, Germany) with a BSD detector. When compared with other detectors, BSD detectors offer better efficiency for in-lens detection with higher surface sensitivity and, consequently, enhanced spatial resolution for resolving pore and grain-level information. The SE detector is used to image particle fusion post-sintering.

\subsection{Numerical estimation of sintering time $t_s$}\label{sec:thermal}
In order to establish the efficacy of the sintering process and, in particular, to estimate the optimal sintering time necessary for the complete part to reach the desired sintering temperature, we employed a numerical model based on a disordered lattice network description \cite{george1988fourier, hrennikoff1941solution, schlangen1997fracture, dawara2022pore}. The basic problem consists of estimating the interior temperature of a porous material (here the consolidated green part) as a function of temperature ramp and hold at the boundaries (here the furnace conditions). The ramp rate was fixe at $5^\circ$ C/min and the temperature at $T_a = 1150^\circ$C, see schematic in Fig.~\ref{schematic}. The lattice network model was then used to estimate the duration of soaking so that the interior of the brick attained uniform temperature to ensure homogeneous sintering. This step is necessary because the interior temperature of the bricks cannot be experimentally monitored during any stage of the process.

\begin{figure}[ht!]
	\centering	
	\includegraphics[width=0.85\textwidth]{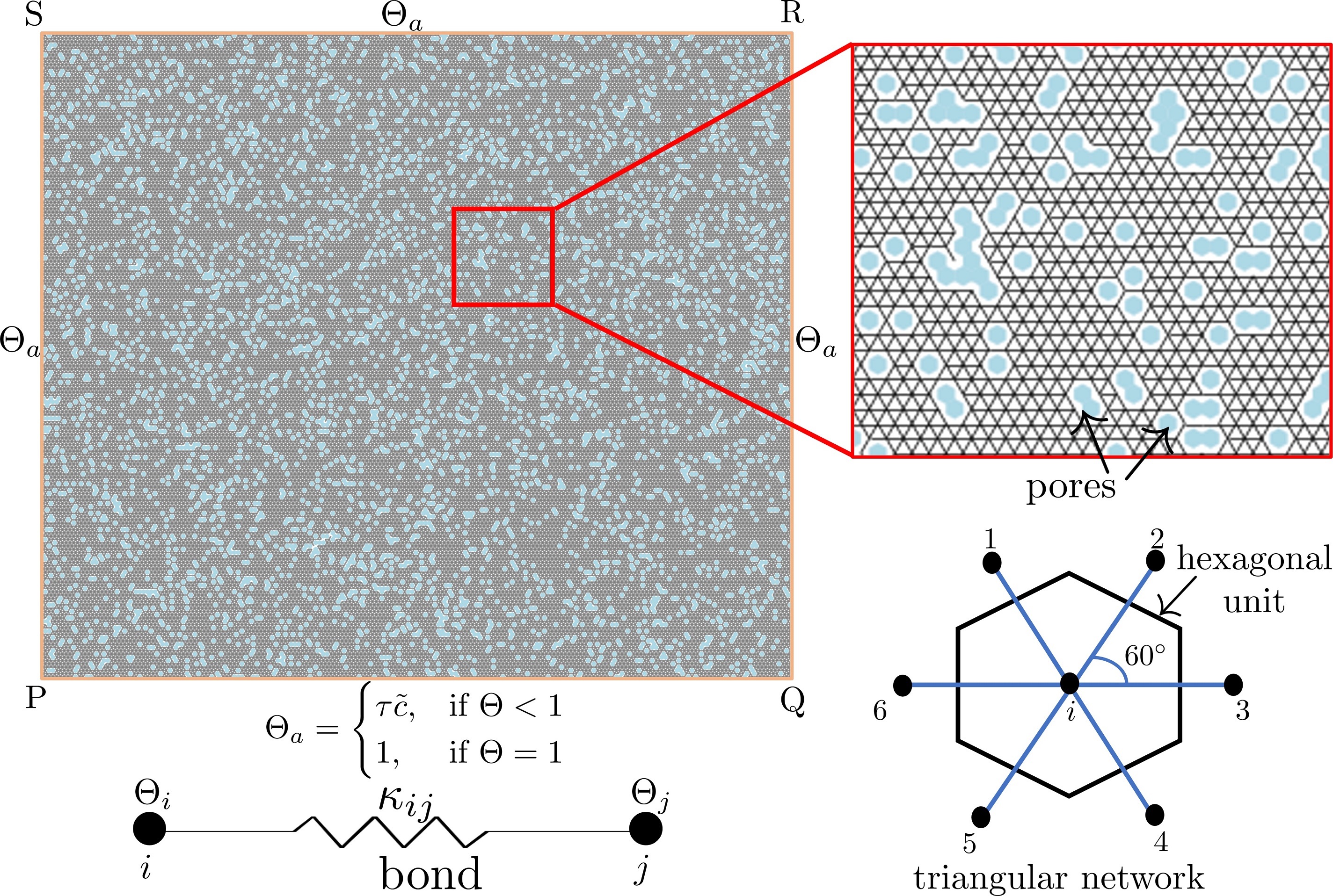}
	\caption{Schematic of lattice network model comprised of triangular lattice nodes and random pore network with porosity $p$. Voronoi polygons were first generated in the triangular lattice with porosity $p$, shown as a zoomed image in the inset. Solid is shown as a regular network of thermal resistors with conductivity $\tilde{\sigma}$ = 2/3 and a temperature difference of ($\theta_j$-$\theta_i$) applied across the bond joining lattice nodes $i,j$; $\tau$, $\Theta$ and $\tilde{c}$ represents non dimensional time, temperature and heating rate, respectively. See text for description.}	
	\label{Simulation_schematic}
\end{figure}

The configuration used for the simulations presented schematically in Fig.~\ref{Simulation_schematic}, and consists of a regular triangular lattice network with unit spacing $a$. The governing heat conduction equation for temperature ($T$) evolution with time ($t$) in an isotropic homogeneous solid with thermal diffusivity $\alpha$ is
\begin{align}
	\frac{\partial T}{\partial t} = \alpha\nabla^2 T            \label{eq:heatconduction}
\end{align}
which, when discretized on this lattice takes the form \cite{martin2000dynamic}
\begin{align}
	\frac{\partial T_i}{\partial t} = \frac{2\alpha}{3a^2}\sum_{j}^{6}(T_j(t) - T_i(t)) \label{eq:heat_spatial}
\end{align}

Thus, we can imagine our solid as a regular network of bonds with diffusivity (unit square bond length) $\kappa_{ij} = 2\alpha/3a^2$  and a temperature difference of $(T_j -T_i)$ applied across it, as described in Fig.~\ref{Simulation_schematic}. To determine the dynamic evolution of temperature, we used forward finite difference time discretization in Eq.~\ref{eq:heat_spatial}, yielding an explicit numerical scheme
\begin{align}
	T_i(t+\Delta t) = T_i(t) + \Delta t \sum \kappa_{ij}\sum_{j}^{6}(T_j(t) - T_i(t)) \label{eq:modelequation}
\end{align}
All material information is described by bond diffusivity $\kappa_{ij}$ between nodes $i,j$. In order to simulate the internal porosity in the green part, we assume that pores are randomly distributed throughout the specimen, which otherwise has uniform diffusivity. Experimentally consistent porosity values $p$ were first obtained from MIP measurements (see Sec.~\ref{charac}), correspondingly nodes were removed in the lattice to generate an equivalent porosity in the network (see inset to Fig.~\ref{Simulation_schematic}). In this way, different realizations of a porous lattice of a given gross porosity $p$ were generated.

At $t = 0$, the temperature at all the nodes was kept equal to room temperature $T_0 = 30^\circ$C. For $t>0$, we assume the nodes on the outer four boundaries of the specimen (orange color, labeled ABCD in schematic) to be at the furnace temperature at all times, which was increased at a constant rate ($c$) from room temperature $T_0$ to $T_a = 1150^\circ$C, and thereafter maintained constant. Equation~\ref{eq:modelequation} was then solved on the porous lattice to determine time needed when the interior to reach the outer furnace temperature $T_a$. This provided the minimum time necessary $t_s$ for the sintering process to occur homogeneously. Data is presented in the form of non-dimensionalized temperature $\Theta = (T - T_0)/(T_a - T_0)$.

\section{Results}\label{results}
We now describe the results of sintering experiments, beginning with optimal soaking time estimation using the lattice network model described in Sec.~\ref{sec:thermal}. We then present results of compression testing experiments and use length measurements during sintering to infer microscale mechanisms operative during sintering. 


\subsection{Porosity and soaking time $t_s$}{\label{simulation}}
Results of the numerical simulations introduced in Sec.~\ref{sec:thermal} are first used to estimate the minimum soaking time $t_S$ necessary for optimal sintering. This is defined as the time taken for the entire network to reach the furnace temperature. However, the porosity $p$ of the green part is an input for the model; we obtain this from MIP measurements of test samples as described in Sec.~\ref{charac}. For bricks sintered at $1150^\circ$C, mercury was infused at 35 kpsi pressure in MIP. The porosity was estimated to be $24.2\%$ and $20.3\%$ for soaking durations of 1 hour and 6 hours, respectively. As an upper bound, we set $p = 25\%$ for the initial network; in the results that follow, temperature $T$ and time $t$ are normalized as
\begin{align}
	\Theta = (T-T_0)/(T_a - T_0)\hspace{1cm}\text{and}\hspace{1cm}  \tau = \alpha t/a^2
\end{align}
Being an explicit scheme, we use a small timestep $\Delta \tau = 0.1$ for ensuring a stable solution. The lattice size was taken to be 200$\times$200, corresponding to the 18$\times$18mm$^2$ dimensions of the green brick; thermal diffusivity of LHS samples was approximated to be 2.65$\times10^{-8}$ m$^2$/s (diffusivities for most simulants are $\sim10^{-8}$ ($m^2$/s)), see \cite{schreiner2016thermophysical, nagihara2022thermal}.

\begin{figure}[ht!]
	\centering	
	\includegraphics[width=0.75\textwidth]{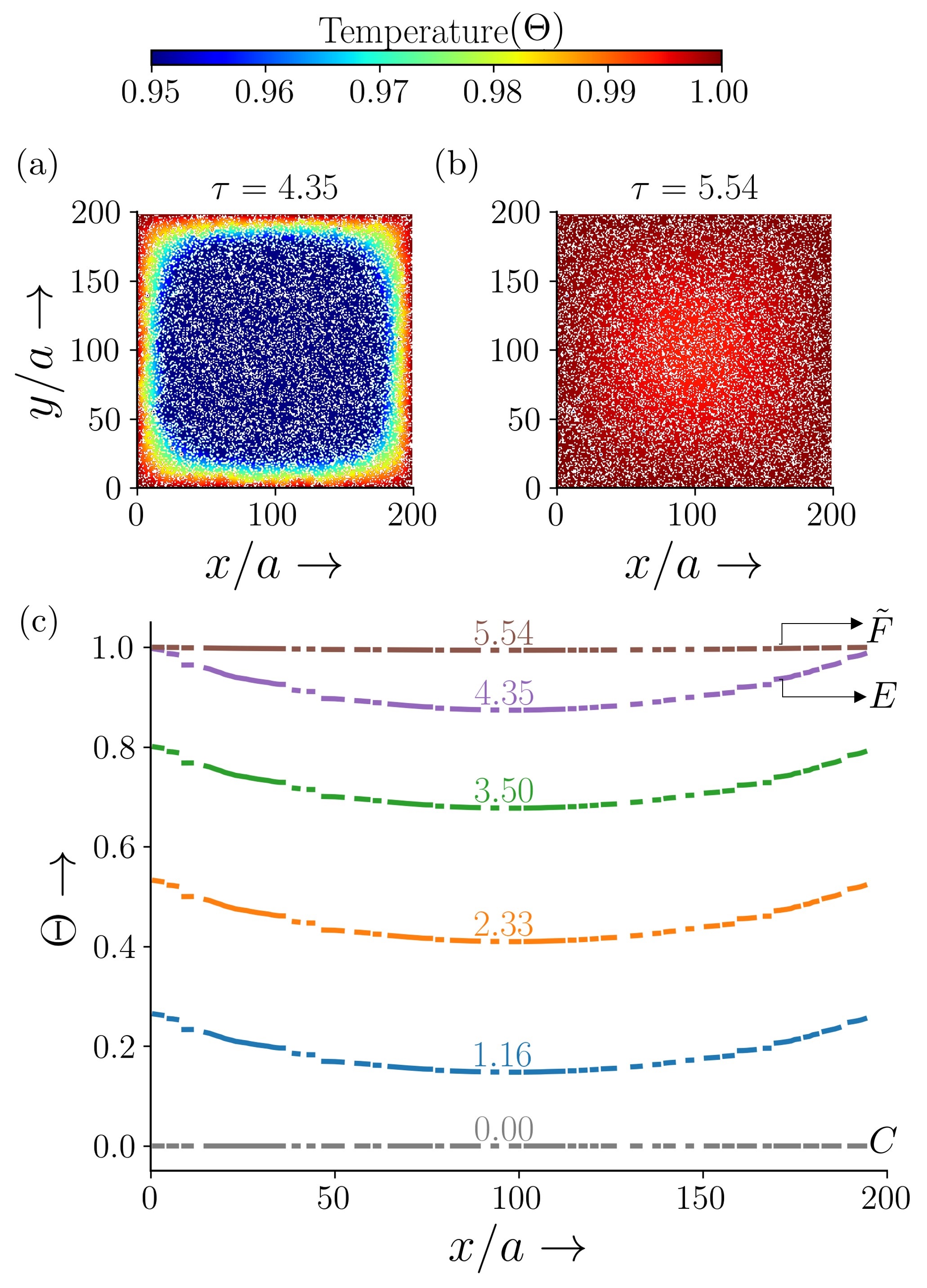}
	\caption{Non-dimensional temperature field in the porous lattice network ($p= 25\%$) when the boundary first reaches the soaking temperature (panel a), and when the interior reaches $\Theta = 1$ (panel b). Panel c shows  temperature distribution ($\theta$) along the horizontal lines at the center for various values of $\tau$ (marked). Thermal conductivity of LHS is 2.65$\times$10$^{-8}$ m$^2$/s, 200$\times$200 lattice size, heat ramping rate 5$^\circ$C/min}
	\label{fig:TM}
\end{figure}

The results of the numerical simulation are summarized in Fig.~\ref{fig:TM}. Panel (a) shows the temperature field in the porous network at time $t_f$ when the furnace first reaches the peak or soaking temperature $\Theta = 1$. A gradient is clearly observed in the field, with $\Theta$ varying by nearly 0.05 from the sample periphery to its interior. As the furnace temperature is held constant at $T_a$ (corresponding to $\Theta = 1$), the interior temperature continues to rise; the time taken for it to reach $T_a$ everywhere inside is set to be the minimum soaking time $t_s$. The corresponding thermal field is shown in panel (b) of  Fig.~\ref{fig:TM}. 

The corresponding temperature evolution along the horizontal midline of the sample is shown in Fig.~\ref{fig:TM}(c) . The heterogeneity of the bricks, majorly pores, causes the fluctuations in the curve. When the heating begins from room temperature $\Theta$ = 0 (point C, in sintering cycle of Fig.~\ref{schematic}), the temperature at all points in the mid of the y-axis is zero. During the temperature ramp, the end temperatures are slowly raised to $\Theta$ = 1 (point E in sintering cycle, Fig.~\ref{schematic}). Even after ramp is complete, the boundaries are maintained at $T_a$, the temperature continues to rise until it becomes uniform inside the sample (point $E$ to $\tilde{F}$).

Based on these simulations, we determine that for the present geometry (cubic, 18 mm side length), a green part requires a total time $t_f+t_s \sim 285$ min to reach uniform peak temperature $T_a$. Given that the furnace takes $t_f = 224$ min to reach $T_a$ from room temperature, the total time for sintering is expected to be atleast $t_s$ $\sim$60 min. This corresponds to the temperature profile curve $\tilde{F}$ in Fig.~\ref{fig:TM}(c). While this value holds specifically for LMS, all simulants used in the present study have similar thermal diffusivity and porosity so that we take 60 minutes as the minimum soaking time for furnace sintering for LHS, LMS and MGS samples.
	
\subsection{Compressive strength of synthetic space bricks}\label{CT}
\begin{figure}[ht!]
	\centering	
	\includegraphics[width=0.7\textwidth]{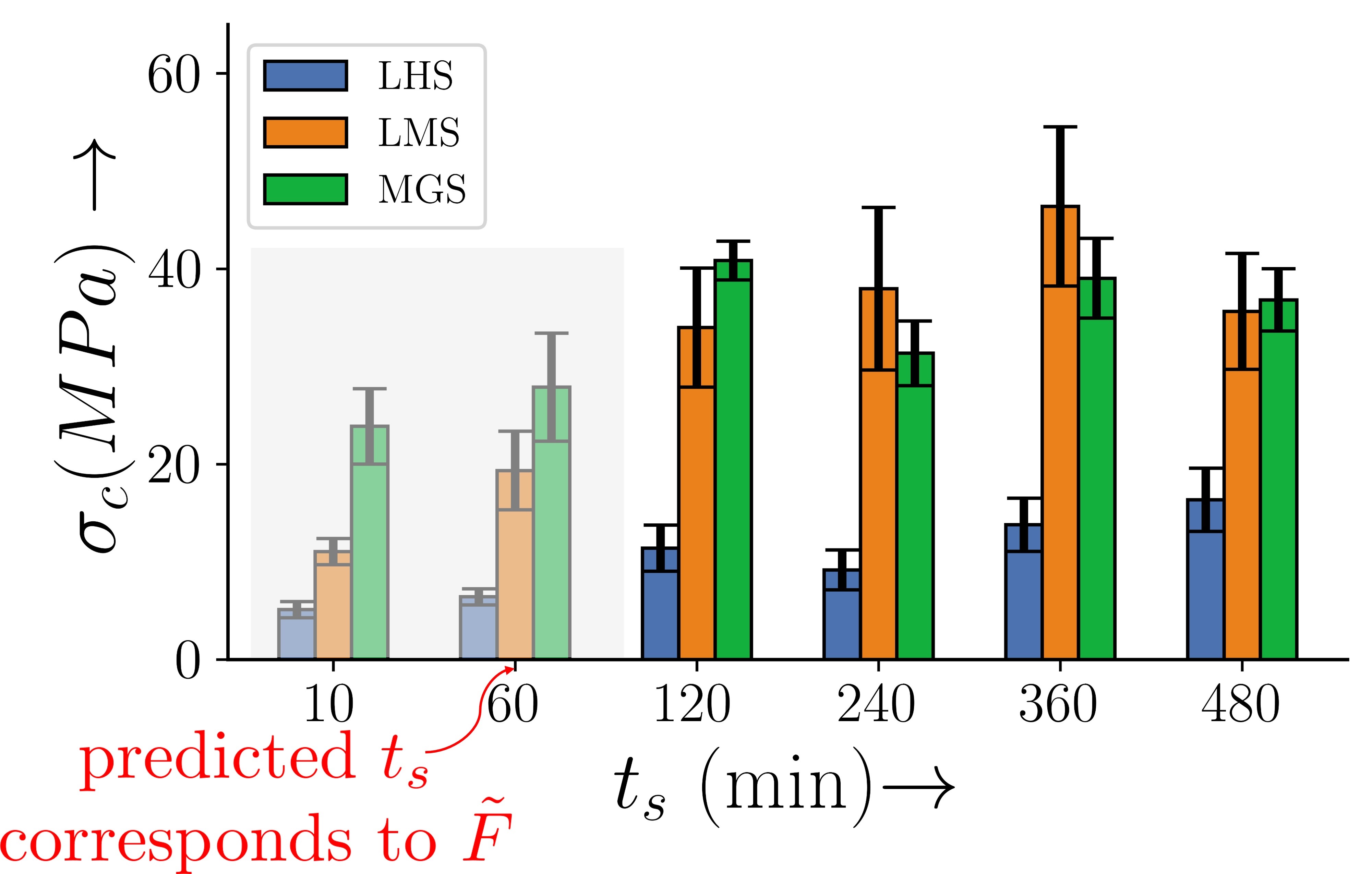}
	\caption{Compressive strength of Synthetic space bricks from LHS (blue), LMS (orange) and MGS (green) simulants as a function of sintering or soaking time. $T_a = 1150^\circ$C.}		
	\label{Strength_1150}
\end{figure}

The compressive strength $\sigma_c$ of sintered samples was measured using unconfined uniaxial compression, as described in Sec.~\ref{charac} and the results summarized in Fig.~\ref{Strength_1150}. For sintering, peak temperature was maintained at $T_a = 1150^\circ$C since that is the maximum temperature at which both LMS and MGS can undergo solid-state sintering, without melting of their constituent components. Data in is compiled based on individual stress-strain curves such as the ones shown in Fig.~S1 of supplementary material. The results are summarized for LHS, LMS and MGS in Fig.~\ref{Strength_1150} in the form of blue, green orange and green bars, respectively. The values here are those obtained over 4 samples with corresponding error bars representing standard deviation. The horizontal axis represents the soaking time $t_s$, and the red arrow marks the cut-off at the minimum soaking time estimated from the previous section, corresponding to curve $\tilde{F}$ in Fig.~\ref{fig:TM}(c).

For $t_s = 10$ min, MGS-based bricks exhibit a mean compressive strength of 23.9 MPa, while LMS and LHS-based bricks showed compressive strengths of 11.0 MPa and 5.1 MPa, respectively. Beyond the estimated minimum $t_s$ of 60 min, significant strength increase was observed for all three cases, with MGS, LMS, and LHS reaching 40.8 MPa, 33.9 MPa, and 11.4 MPa, respectively. This is a clear sign of enhanced sintering due to homogeneous internal temperature. 

However, given that this condition provides only a lower bound for sintering time $t_S$, we evaluated the strength at times $t_S$ upto 480 min, as shown in Fig.~\ref{Strength_1150}. LMS showed the largest compressive strength (exceeding 40 MPa) at $t_S = 360$ min, with a reduction in strength at higher $t_S$. Likewise, the strength of MGS bricks appeared to saturate after $t_S = 240$ min, reaching $\sim 35$ MPa. LMS bricks showed the lowest strength ($< 20$ MPa) throughout the tested $t_S$ range. The reason for the apparent reduction in strength for $t_s > 360$ min for all three cases is not clear, and could be dependent on specific microscopic processes operative in each of the regolith materials. Potential causes include grain growth and microcrack formation due to thermal stress leading to enhanced susceptibility to fracture.

\begin{figure}[ht!]
  \centering
  \includegraphics[width=0.8\textwidth]{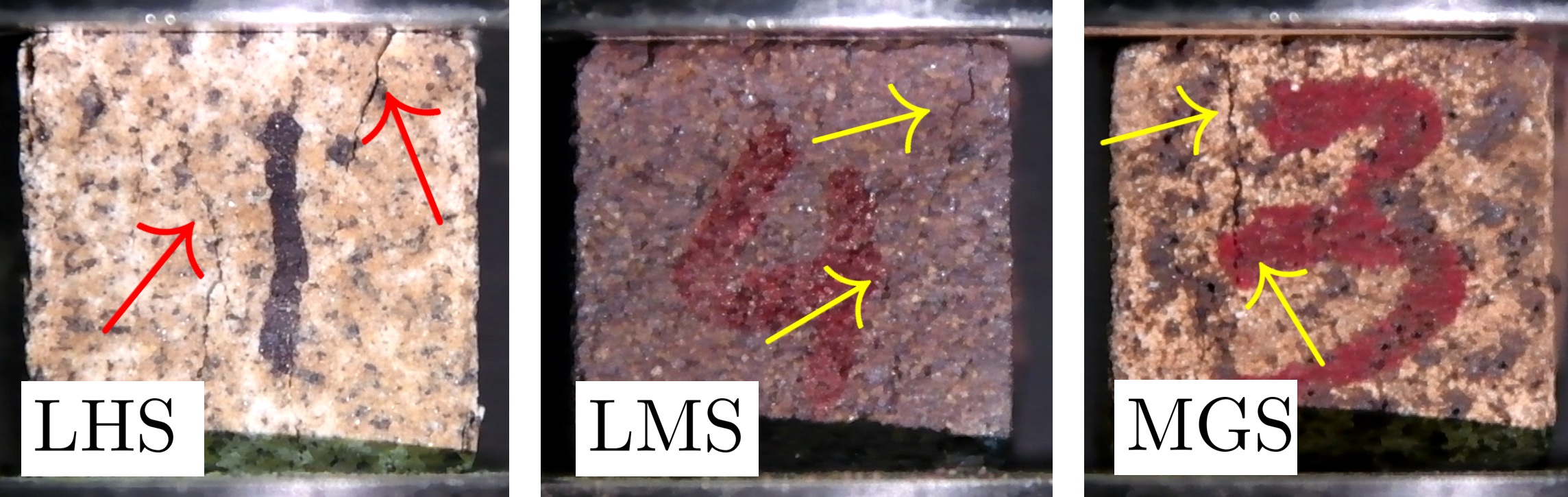}
  \caption{Fracture patterns accompanying failure of synthetic space bricks under unconfined compression. Arrows point to cracks growing in the loading direction. $T_a = 1150^\circ$C, $t_S = 360$ min.}
  \label{fracture}
\end{figure}
      
In each compression test, the compressive strength was evaluated using the maximum force measured during loading, \emph{cf.} Fig.~S1 of supplementary. At this point, the bricks undergo compressive failure, leading to a significant reduction in the force-displacement curve. This failure is mediatead by a number of cracks that are commonly aligned in the loading direction, see Fig.~\ref{fracture}. The three panels in this figure show LHS (left), LMS (middle) and MGS (right) samples, respectively, at the point of failure. All of these samples were sintered for $t_s = 360$ min. In each case, the proliferation of multiple cracks is clear (see at arrows), all nominally aligned with the loading direction. Prior simulations of these fracture patterns have shown that the most likely mechanism for this is pore and microcrack coalescence leading to macroscopic cracks. These then grow in a direction parallel to the compression axis due to the lack of any lateral confining pressure \cite{dawara2022pore}.

\subsection{Post-sintering microstructure}\label{SEM}
\begin{figure}[ht!]
	\centering
	\includegraphics[width=0.85\textwidth]{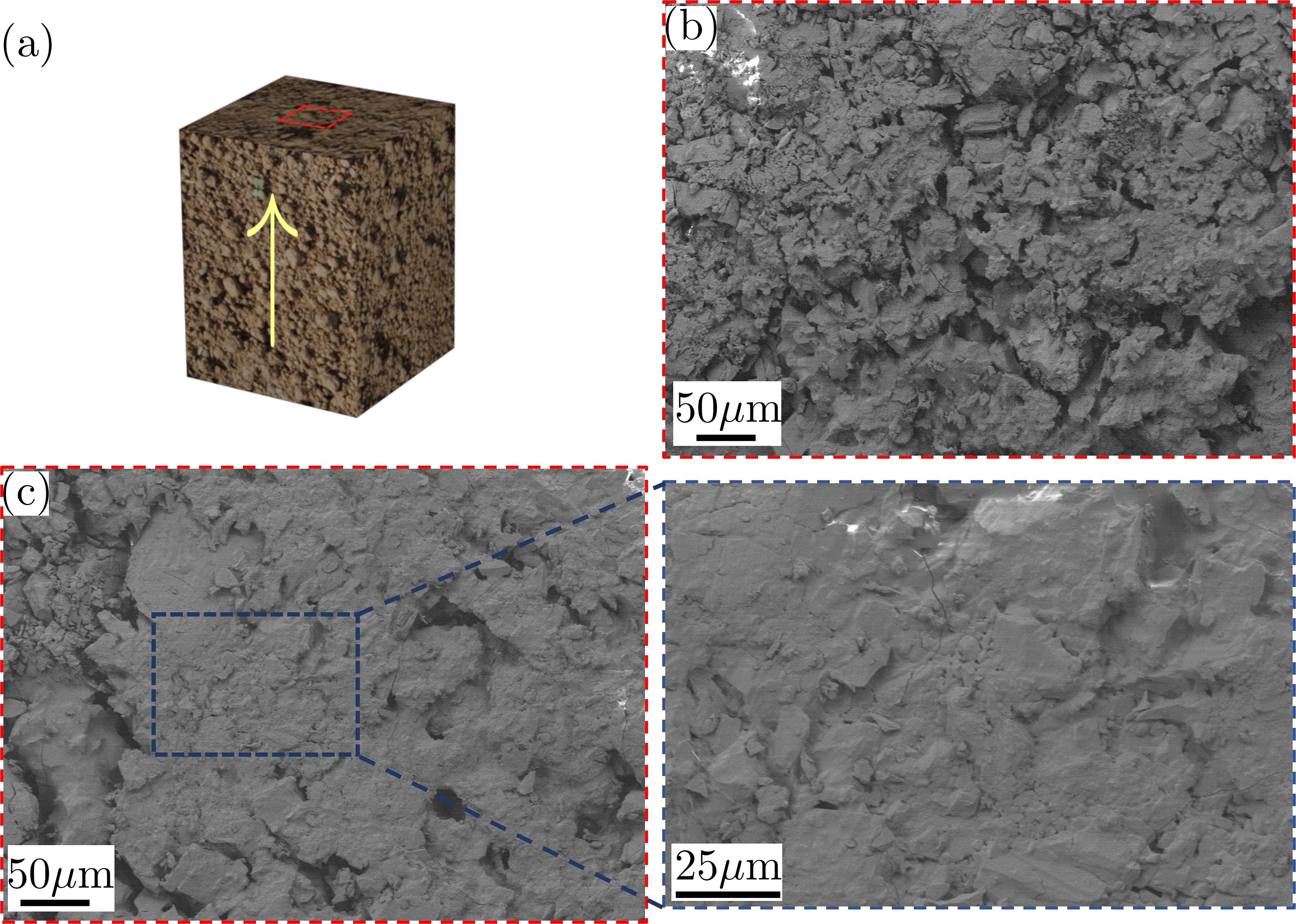}
	\caption{SEM micrographs showing the top layer of LHS-based bricks using BSD detector. (a) Schematic of SEM micrograph location. Panels (b), (c) show micrographs for $t_S$ = 60 min and 480 min, respectively, clearly demonstrating enhanced sintering with $t_S$, and consistent with compressive strength measurements. $T_a = 1150^\circ$C.}
	\label{micrographs}	
\end{figure}

Post-process SEM images of bricks show clear signs of sintering on the microscale, see Fig.~\ref{micrographs}. Panel (a) shows a typical brick and the location at which SEM images are taken; the yellow arrow represents compaction direction. A small section was mechanically removed from the sample surface and gold-coated for imaging. Data for LHS bricks is presented, corresponding images for LMS and MGS appear qualitatively similar. BSD detector was used to obtain these images.

Panels (b) and (c) show the surface after $t_S = 60$ min (minimum $t_S$ estimated) and $t_S = 360$ min, respectively. Temporal progress of the sintering process is clear from these images---initially disparate regolith particles first appear loosely bound after 60 min (panel (b)). They demonstrate significantly enhanced cohesion after 360 min, with a few pores evident between particles (panel (c)). A higher resolution image (inset, blue box) shows that the individuality of particles is clearly no longer discernible at this stage. 

These micrographs show how crucial it is to regulate the sintering parameters to create a coherent structure, which ultimately affects the strength and utility of the bricks. The micrographs for LMS and MGS also show a similar trend, even though the overall strength is quite different in all three cases (\emph{cf.} Fig.~\ref{Strength_1150}). This largely similar microstructure is also perhaps responsible for the similar crack patterns observed during failure of all three regolith-sintered bricks (\emph{cf.} Fig.~\ref{fracture}).

\subsection{Inferring microscale sintering mechanisms}\label{mech}
We now turn our attention to the microscopic mechanisms underlying particle sintering during the soaking stage. Extensive prior studies by the ceramics community dating back to the 1950s have identified four primary candidate mechanisms for the joining of two individual particles---viscous flow, atomic evaporation and condensation, surface diffusion, and lattice diffusion. The underlying assumptions here are that only 2-particle mechanisms are dominant and that particles themselves are approximately spherical.

\begin{figure}[ht!]
	\centering
	\includegraphics[width=0.8\textwidth]{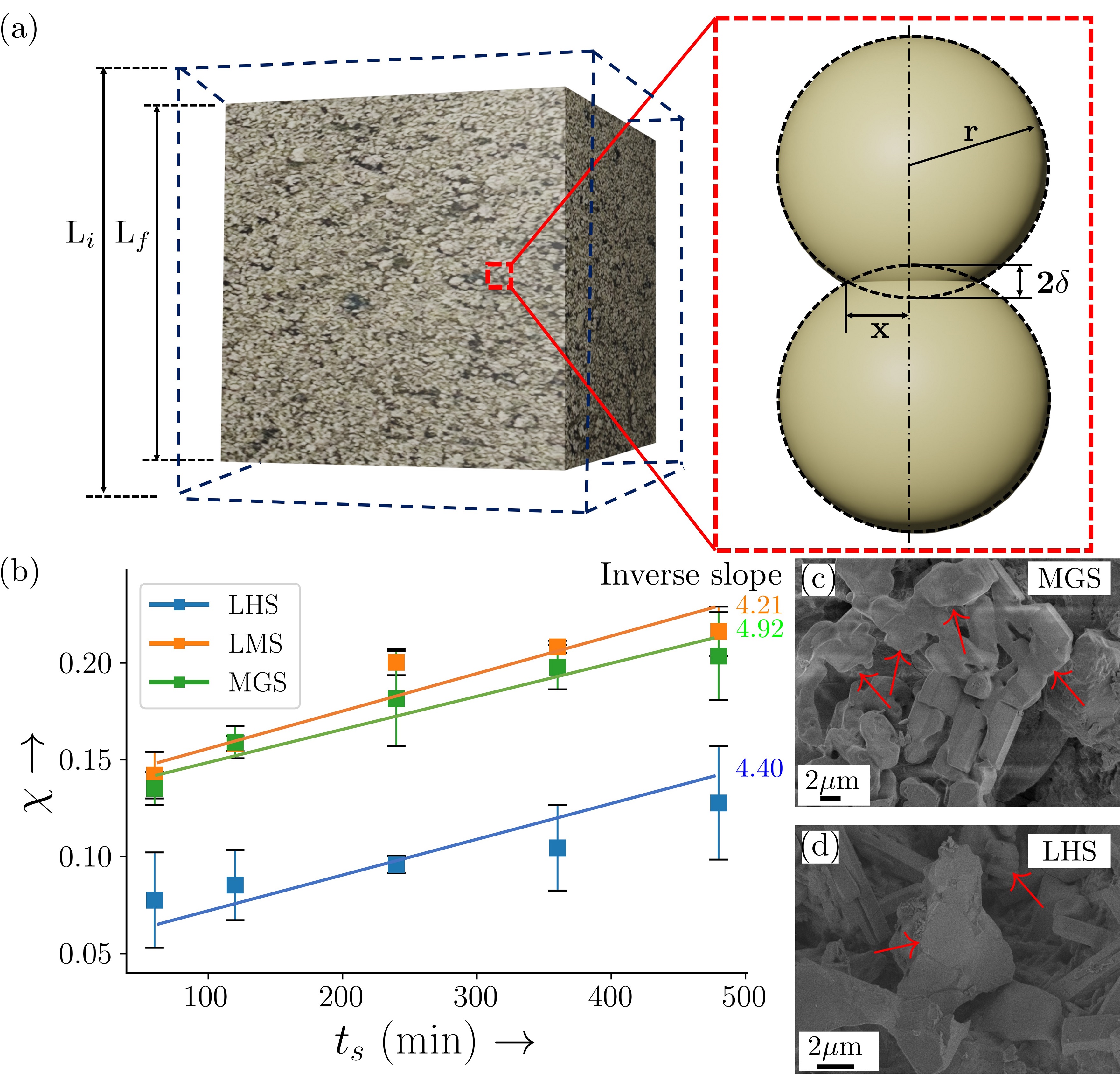}
	\caption{Mechanism of neck growth and particle coalescence in sintered bricks. (a) Schematic representing two-particle model for calculating the growth rate. Panel (b) presenets experimental observations of variation of $\chi$ vs. $t_S$ on a log-log scale, note that $\chi=\dfrac{x}{r} = 2\sqrt{\dfrac{\Delta L}{L}}$. Panels (c) and (d) show SEM micrographs (SE detector) of MGS and LHS, respectively with red arrows indicating particle coalescence. Data in (b) is averaged over 4 samples, $T_a = 1150^\circ$C.}
	\label{mechanism}
\end{figure}

Identifying which of these mechanisms is operative in our bricks is challenging due to the large poly-dispersity in particle size and shape as well as the complex mineral compositions involved. A simple macroscale method for inferring the dominant mechanism(s) involves measuring sample dimensions during the soaking process as a function of $t_S$ \cite{kuczynski2012sintering, kuczynski1949study, johnson1963diffusion, johnson1964grain, johnson1969new, kingery1990study, kang2004sintering, zhang2021spark}. Reduction in dimension is approximately correlated with centre-to-centre distance $\delta$ and neck radius $x$ between particles on the microscale, see Fig.~\ref{mechanism}(a). In general, $x^n\sim  t$, where the exponent $n$ is governed by which mechanism is operative. For viscous flow, evaporation condensation, volume diffusion and surface diffusion, the value $n$ is 2,3,5 and 7, respectively.

Specifically, 
\begin{equation}
  \label{eqn:exponent}
  \Big(\dfrac{x}{r}\Big)^n= \text{At} \implies \log\Big(\dfrac{x}{r}\Big) = \log{(\chi)}= \dfrac{1}{n}\log(A) + m\log(t) 
\end{equation}
where $A$ is a material-dependent constant. For the bricks, $x/r$ is obtained approximately by measuring the percentage shrinkage in linear dimensions from the green to the brown part \cite{johnson1969new}.
\begin{equation}
  \label{eqn:shrinkage}
  \text{shrinkage} = \dfrac{L_i -L_f}{L_i} = \dfrac{\Delta L}{L_i} = \dfrac{\delta}{r} \approx \dfrac{x^2}{4 r^2} = \dfrac{\chi^2}{4}
\end{equation}	
where $2\delta$ is change in centre-to-centre distance between two spherical particles under coalescence, and $r$ is the particle size.

In order to finally evaluate the exponent $n$, dimensions of the cubic green and brown parts were measured using a vernier caliper at various locations and averaged to get a mean linear dimension; the value of $\Delta L/L_i$ then provided an estimate of $\chi$ from Eq.~\ref{eqn:shrinkage}. The corresponding log($\chi$) vs. log($t_s$) plot is shown in Fig.~\ref{mechanism}(b). The relationship appears to be nearly linear, with slope equal to the inverse exponent $m = 1/n$, see Eq.~\ref{eqn:exponent}. We have only used samples for which $t_s > 60$ min in accordance with the estimate obtained in Sec.~\ref{simulation}. The error bars represent standard deviations over 4 successive measurements.

The approximated slopes from the curves in Fig.~\ref{mechanism}(b) correspond to $n =$ 4.4, 4.2, and 4.9 for LHS, LMS, and MGS, respectively, sintered at 1150$^\circ$C. These values strongly suggest that volume diffusion ($n=5$) is the predominant mechanism for sintering on the microscale. Corresponding SEM images (SE detector) of MGS and LHS are shown in Fig.~\ref{mechanism}(c) and (d), respectively. The coalescence of individual regolith particles appears quite clear (at arrows). The extent to which other mechanisms (e.g., viscous flow) are operative remains uncertain based on these investigations and certainly warrants further study.

\section{Discussion and Summary}
\label{sec:conclusions}

Based on our investigations, it is clear that a significant difference in compressive strength (unconfined) is to be expected between various regolith simulants---LMS and MGS are comparable ($\sim 40$ MPa) while that of LHS is nearly $60\%$ lower. It is entirely possible that the use of higher sintering temperature may significantly alter this picture since fundamental chemical changes are expected, given the soil composition. In fact, exploiting the presence of increased glass basalt content in LHS appears to be a direct route for enhancing sintering strength that we are presently pursuing.

As fundamental building blocks for habitat applications, the minimum strength needed for sustaining their self weight is around 3 MPa (lunar) and 6 MPa (martian), based on the lower gravity on the surface of the moon or mars, respectively. The synthetic space bricks reported in our work have significantly larger strength, making them more than suitable for these applications, even if only partially sintered (\emph{cf.} Fig.~\ref{Strength_1150}). As far as deployability is concerned, the sintering technique is ideally suited to \emph{in situ} resource utilization (ISRU) on extra terrestrial habitats since the process can be scaled and completely automated. We believe that this large strength and process scalability are fundamental advantages of the sintering process, over other routes that have been proposed in the literature, see also Table.~\ref{sintering_LR}.

For both the lunar and martian bricks, ultimate failure (under unconfined compression) occurs via the propagation of multiple axis-aligned cracks. A typical stress-strain curve for these bricks (se Fig.~S1, supplementary material) shows that they are essentially brittle with little plastic flow on the macroscale. The growth of cracks occurs due to local stress concentration---at either large pores or pre-existing microcracks---that can lead to catastrophic failure at a size-dependent critical load. This behaviour was also observed to be somewhat anisotropic, being different in the compaction direction (initial green part production) vis-\'a-vis the transverse directions. In general, pores, inclusions, and grain boundaries are examples of microstructural flaws that may inherently operate as stress concentrators, and provide locations for crack initiation and propagation.

Based on our results, it is to be expected that the peak temperature and soaking time are the two primary contributors to final part strength. While the increase in strength with $t_S$ is to be expected---longer soaking time leads to better particle coalescence and hence, enhanced sintering---the reduction in strength for $t_S > 360$ min across all three materials is noteworthy. This is quite likely due to the occurrence of large internal pores, either driven by vacancy diffusion in the bulk or by thermal stresses. However, little evidence of this was found in the compression test failure mechanisms and this thus warrants further investigation. As mentioned in the text, the occurrence of high glass basalt content (melting $\sim 1170^\circ$C) in LHS is likely to increase the strength in samples sintered at higher temperature; we are presently actively investigating this possibility.

In the context of sintering mechanisms, our results, based on classical analyses commonplace in the ceramics community, indicate that volume diffusion is the primary driver of particle coalescence. This is clear based on the macroscopic measurements presented in Fig.~\ref{mechanism}, yet direct evidence of these mechanisms is fundamentally challenging to obtain. The possibility of directly observing individual particle coalescence at the grain level is complicated by both the length-scales and high temperatures involved, not to mention the complete lack of axisymmetry in the particles themselves. We believe that coarse-grained granular dynamics simulations could be profitably employed to address this question in more detail.

In summary, our work proposes a procedural protocol for fabricating synthetic space bricks via sintering in extra-terrestrial settings using \emph{in situ} resource utilization. We have demonstrated the potential for using both lunar and martian regolith simulants, with significant compressive strengths of final sintered bricks. Process parameters were estimated using a numerical model to evaluate time needed for uniform temperature in the interior of porous green parts produced using a polymer binder. Based on these results, compressive strength measurements were performed as a function of the sintering time at a fixed temperature. Our results show that strengths of $>40$ MPa are achievable with both lunar (LMS) and martian (MGS) simulants. Strength resulting from particle coalescence was confirmed using both compression testing and SEM imaging of consolidated samples. Based on linear shrinkage, we inferred the primary microscopic sintering mechanism to be volume diffusion in the bulk of individual particles in all three simulants. Based on our results, the potential for using sintering to consolidate regolith into bricks for structural applications has been clearly demonstrated.

\section*{Acknowledgments}
The authors acknowledge Mr Bhupendra Chand and Prof Tejas Murthy, Civil Engineering, IISc for extending the MIP facilities for conducting porosity analysis.

	\clearpage	
	{\footnotesize
		\bibliography{bibfile}}
	\bibliographystyle{unsrt}	
\end{document}